\documentclass[12pt]{iopart}

\usepackage{fontspec}
\usepackage{xeCJK}
\usepackage{deluxetable}
\usepackage{url}
\usepackage{graphicx}

\expandafter\let\csname equation*\endcsname\relax
\expandafter\let\csname endequation*\endcsname\relax
\usepackage{amsmath}
\usepackage{amssymb}

\begin{document}

%
%
%

\title[]{From Jets to Failed Supernovae: Morphologies and Gravitational-Wave Signatures in Two-Dimensional Magnetorotational Core-Collapse Supernovae}

\author{Kuo-Chuan Pan (潘國全)$^{1,2, 3}$, Yi-Fang Li (李宜舫)$^{1,2}$}
\address{$^1$Department of Physics, National Tsing Hua University, Hsinchu, Taiwan}
\address{$^2$Institute of Astronomy, National Tsing Hua University, Hsinchu, Taiwan}
\address{$^3$Department of Physics and Astronomy, North Carolina State University, Raleigh, NC 27695, USA }

\ead{kuochuan.pan@gapp.nthu.edu.tw}

\vspace{10pt}
\begin{indented}
\item[]December 2025
\end{indented}

\begin{abstract}
Magnetized and rotating core-collapse supernovae (CCSNe) are promising candidates for producing long gamma-ray bursts and hypernovae. In this project, we present 34 two-dimensional magnetized core-collapse supernova simulations with self-consistent neutrino transport, systematically exploring the parameter space of initial magnetic field strengths ($B_0 = 0$--$3.5 \times 10^{12}$~G) and rotation rates ($\Omega_0 = 0$--$0.5$~rad~s$^{-1}$) for a 40~$M_\odot$ progenitor. Our simulations reveal four distinct explosion morphologies: failed explosions leading to black hole formation, monopolar jet explosions, bipolar jet explosions, and neutrino-driven explosions. We find that the 40 $M_\odot$ progenitor model failed to explode without magnetic fields in two dimensions, even with rapid rotation. The non-rotating models require strong seed magnetic fields ($B_0 \gtrsim 1.5 \times 10^{12}$~G) to launch magnetically driven explosions, while the introduction of rotation substantially lowers this threshold. The explosion timescale decreases systematically with both increasing magnetic field strength and rotation rate, ranging from $>500$~ms in marginally successful models to $<150$~ms in strongly magnetized, rapidly rotating systems. Diagnostic explosion energies in the most extreme models approach $\sim 10^{51}$~erg within 250~ms and continue growing in time, making them potential hypernovae and long gamma-ray burst progenitors. Finally, we analyze the gravitational wave signatures associated with each morphology and find that the gravitational wave frequencies mainly depend on the rotation rates but are less sensitive to the magnetic field strengths and explosion morphologies. However, the gravitational wave amplitudes strongly depend on the explosion morphologies and magnetic fields, making searches for gravitational waves from magnetorotational core-collapse supernovae more challenging.
\end{abstract}

%
%
\submitto{\CQG}
%
%

\section{Introduction}

Core-collapse supernovae (CCSNe) mark the catastrophic end of massive stars,
serving as the cosmic engines driving the formation of neutron stars (NSs) and stellar-mass black holes (BHs) \cite{1990RvMP...62..801B,2007PhR...442...38J,2009ARA&A..47...63S,2013RvMP...85..245B,2020LRCA....6....3M}.
These events are recognized as critical multi-messenger sources, emitting electromagnetic radiation, bursts of MeV neutrinos, and gravitational waves (GWs) \cite{1987PhRvL..58.1490H,1987PhRvL..58.1494B,2006RPPh...69..971K,2012ARNPS..62..407J,2016ARNPS..66..341J,2016NCimR..39....1M}.
The detection of GWs from a nearby CCSN remains a significant objective, offering crucial, direct insights into the explosion mechanism and the fundamental properties of the collapsing core \cite{2002ApJ...565..430F,2009CQGra..26f3001O,2013CRPhy..14..318K,2020PhRvD.101h4002A}.
While the standard neutrino-driven mechanism adequately explains the majority of observed CCSNe (with typical energies $\sim 10^{51} \text{erg}$),
it fails to account for highly energetic transients, such as Type Ic-BL supernovae and hypernovae,
which possess kinetic energies reaching $\sim 10^{52} \text{ erg}$ and often feature relativistic outflows \cite{1998Natur.395..672I,2003ApJ...599L..95M,2006ARA&A..44..507W,2016ApJ...832..108M}.
These extreme events, frequently associated with long $\gamma$-ray bursts (GRBs), necessitate a more potent energy mechanism,
specifically the magnetorotational mechanism, which relies on the rapid rotation of the progenitor core
and the generation of magnetar-strength magnetic fields ($\gtrsim 10^{15} \text{ G}$) to launch powerful bipolar jets \cite{1970ApJ...161..541L,1992ApJ...392L...9D,1999ApJ...524L.107K,2007ApJ...664..416B,2017ARA&A..55..261K}.

The implementation of the magnetorotational mechanism in numerical simulations faces fundamental physical
and computational challenges stemming from uncertainties in the initial conditions.
The initial ``seed" magnetic field ($B_0$) in progenitor cores is typically weak ($\ll 10^{15} \text{ G}$),
requiring robust amplification via plasma instabilities such as the Magnetorotational Instability (MRI)
or dynamo action to reach dynamical relevance shortly after core bounce \cite{1991ApJ...376..214B,2003ApJ...584..954A,2015MNRAS.450.2153G,2015Natur.528..376M,2016MNRAS.460.3316R}.
A major computational bottleneck is the extreme resolution required to capture the fastest growing mode (FGM) of the MRI,
demanding linear grid spacings of $\lesssim 50-100 \text{ m}$ in global simulations \cite{2009A&A...498..241O,2013ApJ...770L..19S,2016ApJ...817..153S}.
To circumvent this prohibitive computational cost, many studies, including earlier work on magnetorotational explosions,
employed unphysically high initial seed magnetic fields ($B_0 \geq 10^{12} \text{ G}$) to mimic the post-amplification state \cite{2007ApJ...664..416B,2012ApJ...750L..22W,2015ApJ...810..109N}.
Similarly, establishing realistic fast-rotating progenitor profiles ($\Omega_0$) remains an active area of stellar evolution research,
although recent studies incorporating binary stellar evolution suggest mass transfer as a viable pathway to produce the necessary rapid rotation rates \cite{2005ApJ...626..350H,2013ApJ...764..166D,2019MNRAS.485.3661F,2020ApJ...901..114A}.

Full three-dimensional (3D) simulations are the gold standard for capturing the non-axisymmetric dynamics inherent to these explosions.
In particular, the MHD kink instability ($m=1$) can fundamentally alter jet morphology and propagation \cite{2014ApJ...785L..29M},
although 3D studies have also shown that jets can remain stable when magnetic fields are strong and propagation speeds are high \cite{2018ApJ...864..171M,2021MNRAS.503.4942O}.
Beyond the magnetorotational regime, three-dimensional MHD simulations have demonstrated that magnetic fields can also play a supportive role in neutrino-driven explosions, either through small-scale dynamo action amplifying fields to near equipartition in the gain region \cite{2020MNRAS.498L.109M,2022MNRAS.516.1752M} or via the turbulent $\alpha$-effect that fosters faster shock revival in rotating models \cite{2024MNRAS.528L..96M}.
The gravitational wave signatures of magnetized, rotating 3D models have also been explored \cite{2024MNRAS.532.4326P}.
Three-dimensional GRMHD simulations of magnetorotational core collapse have been carried out by a number of groups \cite{2014ApJ...785L..29M, 2018ApJ...864..171M, 2020ApJ...896..102K, 2021MNRAS.503.4942O, 2024MNRAS.531.3732S}, though the immense computational expense has generally limited either the breadth of the parameter space or the simulation duration. The recent development of GPU-accelerated codes such as \texttt{GRaM-X} \cite{2023CQGra..40t5009S} is now enabling more systematic parameter studies \cite{2025arXiv250411537S, 2026MNRAS.546ag056S}, but comprehensive coverage of the progenitor landscape remains challenging.
Axisymmetric two-dimensional (2D) magnetohydrodynamic simulations therefore remain an indispensable complement,
particularly for large parameter surveys with detailed neutrino transport that efficiently map the landscape of explosion outcomes, energy yields,
and multi-messenger signals across initial magnetic field and rotational properties \cite{2009ApJ...691.1360T,2020MNRAS.492.4613O}.

In this paper, we present a comprehensive parameter-space study of magnetized, rotating CCSNe using 34 two-dimensional MHD simulations.
We systematically explore a wide range of initial magnetic field strengths ($B_0 = 0- 3.5 \times 10^{12} \text{ G}$)
and rotation rates ($\Omega_0 = 0-0.5 \text{ rad s}^{-1}$) using the s40, $40 M_{\odot}$ progenitor model from \cite{2007PhR...442..269W}, a massive progenitor that is expected to fail as a neutrino-driven supernova \cite{2016ApJ...821...38S}.
The paper is organized as follows.
In Section~\ref{sec:methods}, we describe the numerical methods and the initial conditions used for our simulations.
In Section~\ref{sec:overview}, we present an overview of all our simulation models with different magnetic field strengths and rotational speeds.
In Section~\ref{sec:explosion}, we analyze the morphology dependence in our parameter space and discuss the impact on ejecta dynamics.
The corresponding analysis on gravitational wave emissions is presented in Section~\ref{sec:gw}.
Finally, we summarize our results and conclude in Section~\ref{sec:summary}.

\section{Methods} \label{sec:methods}

Our magnetized CCSN simulations employ a customized version of the Eulerian multidimensional hydrodynamics code FLASH \cite{2000ApJS..131..273F},
following a setup similar to \cite{2016ApJ...817...72P, 2018ApJ...857...13P, 2019ApJ...878...13P, 2019JPhG...46a4001P, 2021ApJ...914..140P, 2024ApJ...964...23W}; we summarize the key points below for completeness.

The ideal magnetohydrodynamic (MHD) equations are solved in two-and-a-half-dimensional cylindrical coordinates ($\rho, z, \phi$) using FLASH's directionally unsplit staggered mesh (USM) algorithm \cite{2009JCoPh.228..952L,2013JCoPh.243..269L}.
Self-gravity is treated via the improved multipole Poisson solver of \cite{2013ApJ...778..181C} with a maximum multipole order of $l_{\rm max}=16$,
and the monopole term is replaced by an effective general relativistic (GR) potential correction (Case A) following \cite{2006A&A...445..273M}.

For neutrino transport, we adopt the Isotropic Diffusion Source Approximation (IDSA) \cite{2009ApJ...698.1174L} for electron-flavor neutrinos, using twenty energy bins spaced logarithmically from 3 MeV to 300 MeV.
Standard Bruenn weak interaction rates \cite{1985ApJS...58..771B} are employed, with neutrino-electron scattering (NES) effectively included only during the collapse phase via a parametrized deleptonization method \cite{2005ApJ...633.1042L}.
Heavy-lepton neutrinos ($\mu$ and $\tau$ flavors and their anti-particles) are treated with the leakage scheme of \cite{2003MNRAS.342..673R}.

The nuclear equation of state is the Steiner, Fischer, and Hempel (SFHo) EoS \cite{2013ApJ...774...17S}, which incorporates density-dependent relativistic mean-field interactions \cite{2010PhRvC..81a5803T} and is tuned to fit neutron star radius observations \cite{2013ApJ...774...17S}.

The simulation domain covers the inner $10^4$~km of the progenitor, spanning the full $180^\circ$ meridional plane in 2D cylindrical coordinates.
Nine levels of adaptive mesh refinement (AMR) are applied on a grid of $5\times 10$ basic blocks in the $r$- and $z$-directions, yielding a minimum cell width of 0.488~km.
A minimum effective angular resolution of $0.2^\circ$--$0.4^\circ$ is imposed at large radii to reduce computational cost, while the maximum refinement level ($\Delta r=\Delta z=0.488$~km) is enforced within the shocked region to accurately capture post-bounce convective motions and mitigate numerical artifacts at grid interfaces.

The initial condition is the solar-metallicity, 40~$M_\odot$ progenitor model of \cite{2007PhR...442..269W}. To ensure the divergence-free constraint ($\nabla \cdot \mathbf{B} = 0$), the magnetic field is initialized through the magnetic vector potential $\mathbf{A}$. We prescribe a purely poloidal field by setting $A_\rho=A_z=0$ and defining the toroidal component as:
\begin{equation}
A_\phi(r) = \frac{B_0}{2} \rho \frac{r_0^3}{r^3 + r_0^3},
\end{equation}
where $B_0$ determines the magnetic field strength at the beginning of the core collapse, $\rho$ is the cylindrical radius, $r = \sqrt{\rho^2 + z^2}$ is the spherical distance from the coordinate center, and $r_0=1000$~km is the characteristic length scale of the magnetic field.
The parameter $B_0$ sets the initial magnetic field strength at the onset of core collapse and is expressed in FLASH code units, which adopt Heaviside-Lorentz conventions such that $B_\mathrm{CGS} = \sqrt{4\pi}\, B_\mathrm{code}$.

A parametrized differential rotation profile is imposed:
\begin{equation}
\Omega(r) = \frac{\Omega_0}{1+ \left(r/A \right)^2},
\end{equation}
where $\Omega_0$ represents the central angular velocity and $A=1000$~km is the differential rotation length scale parameter \cite{1985A&A...147..161E}.
We explore a wide range of initial conditions, varying $B_0$ from $0$ to $10^{12}$ ($\sim 3.5 \times 10^{12}$~G; in logarithmic increments) and $\Omega_0$ from $0$ to $0.5~\text{rad~s}^{-1}$.
Each simulation begins at the onset of core collapse and evolves self-consistently until either a successful explosion develops (maximum shock radius $\approx 1000~\text{km}$) or a failed supernova results in black hole formation.

Black hole formation is identified by a sudden increment in central density, corresponding to the moment when matter crosses the apparent horizon.
Because we employ an effective GR potential rather than full general relativity,
the black hole formation times reported here should be considered approximate
(see \cite{2011ApJ...730...70O} for a systematic study of BH formation in failing CCSNe using the same $s$40 progenitor).

Several diagnostic quantities are defined for the analysis that follows.
The proto-neutron star mass ($M_{\rm pns}$) in Table~\ref{tab:simulations} is the \textit{baryonic} mass
enclosed within a density threshold of $\rho > 10^{11}~\text{g~cm}^{-3}$.
The diagnostic explosion energy ($E_{\rm dia}$) is the volume integral of the total specific energy (kinetic $+$ internal $+$ gravitational) over all material with positive total specific energy. Note that the specific internal energy used here has subtracted a reference energy, corresponding to the minimum internal energy in the EOS table with the same density and electron traction \cite{2021ApJ...914..140P}. 
Gravitational wave signals are extracted using the standard Newtonian quadrupole formula
in the slow-motion, weak-field approximation \cite{2009CQGra..26f3001O, 2018ApJ...857...13P},
computing the time-varying mass quadrupole moment from the 2D density and velocity distributions.
In axisymmetry, only the $m=0$ component of the quadrupole radiation is captured;
non-axisymmetric contributions (e.g., from spiral SASI modes or bar-mode instabilities) are absent by construction.

Based on the simulation outcomes, we classify the explosion morphologies into four categories:
(1)~\textit{Failed explosions} (BH): the shock stagnates and the central density undergoes sudden collapse;
(2)~\textit{Monopolar jet explosions} (MP): shock expansion exceeds $500~\text{km}$ along only one polar direction;
(3)~\textit{Bipolar jet explosions} (BP): shock expansion exceeds $500~\text{km}$ along both poles with jet opening angles $\lesssim 45^\circ$;
(4)~\textit{Neutrino-driven explosions} (ND): more isotropic shock expansion with broad opening angles ($\gtrsim 45^\circ$), driven primarily by neutrino heating.
Model B9\_Omg04, labeled ``X'' in Table~\ref{tab:simulations}, represents an early crashed case
that does not fit cleanly into any single category.

An important caveat concerns the grid resolution: our minimum cell width ($\Delta x_{\rm min} \approx 0.488~\text{km}$)
is insufficient to resolve the fastest-growing mode of the magnetorotational instability (MRI)
in most of the proto-neutron star interior.
The MRI wavelength $\lambda_{\rm MRI} \sim 2\pi v_A / \Omega$ ranges from $\lesssim$ a few km
for the strongest post-bounce fields ($B \sim 10^{15}~\text{G}$)
to tens of meters for weaker fields, both of which are at or below our grid scale
(see also \cite{2016ApJ...817..153S} for a detailed resolution study).
Our simulations therefore capture large-scale magnetic field amplification via compression and rotational winding,
but not the fully developed MRI-driven turbulent cascade, albeit the 0.488~km resolutions in all the shocked regions are one of the highest resolutions setups in global CCSN simulations with neutrino transport.  

Table~\ref{tab:simulations} provides a comprehensive summary of all 34 simulation models, including the initial parameters, final morphology classifications, explosion timescales, proto-neutron star properties, and energetics for each model.

\begin{deluxetable}{lccccccccc}
\tabletypesize{\footnotesize}
\tablecolumns{10}
\tablewidth{0pt}
\tablecaption{ Simulations \label{tab:simulations}}
\tablehead{
\colhead{Name} & \colhead{$B_0$} & \colhead{$\Omega_0$}  & \colhead{Morph.} & \colhead{$t_{\rm end}$}  &  \colhead{$R_{\rm sh,end}$}& \colhead{$M_{\rm pns,end}$} & \colhead{$E_{\rm mag}$} & \colhead{$E_{\rm kin}$} & \colhead{$E_{\rm dia}$}\\
\colhead{} & \colhead{(Code)} & \colhead{(rad/s)} & \colhead{} & \colhead{(ms)} & \colhead{(km)} & \colhead{($M_\odot$)} & \colhead{(erg)}  & \colhead{(erg)}  & \colhead{(erg)}}
\startdata
B0\_Omg0 & $0$ & 0 & BH & 642.2 & 40.4 & 2.49 & 0 & $4.19 \times 10^{48}$ & -- \\
B0\_Omg01 & $0$ & 0.1 & BH & 687.3 & 47.6 & 2.52 & 0 & $4.77 \times 10^{48}$ & -- \\
B0\_Omg02 & $0$ & 0.2 & BH & 769.0 & 41.6 & 2.57 & 0 & $5.94 \times 10^{48}$ & -- \\
B0\_Omg03 & $0$ & 0.3 & BH & 920.9 & 48.7 & 2.65 & 0 & $9.06 \times 10^{48}$ & -- \\
B0\_Omg04 & $0$ & 0.4 & BH & 1115.2 & 63.3 & 2.76 & 0 & $1.31 \times 10^{49}$ & -- \\
B0\_Omg05 & $0$ & 0.5 & BH & 1057.5 & 66.8 & 2.73 & 0 & $1.32 \times 10^{49}$ & -- \\
B9\_Omg0 & $10^{9}$ & 0 & BH & 638.7 & 44.9 & 2.49 & $3.38 \times 10^{42}$ & $4.99 \times 10^{48}$ & -- \\
B9\_Omg01 & $10^{9}$ & 0.1 & MP & 306.0 & 725.5 & 2.21 & $4.32 \times 10^{48}$ & $1.66 \times 10^{49}$ & $4.97 \times 10^{49}$ \\
B9\_Omg02 & $10^{9}$ & 0.2 & BP & 277.1 & 844.7 & 2.15 & $4.65 \times 10^{49}$ & $1.59 \times 10^{50}$ & $2.92 \times 10^{50}$ \\
B9\_Omg03 & $10^{9}$ & 0.3 & MP & 115.6 & 198.4 & 1.87 & $7.28 \times 10^{47}$ & $1.10 \times 10^{49}$ & -- \\
B9\_Omg04 & $10^{9}$ & 0.4 & X & 138.1 & 209.4 & 1.92 & $8.78 \times 10^{47}$ & $1.73 \times 10^{49}$ & $3.10 \times 10^{49}$ \\
B9\_Omg05 & $10^{9}$ & 0.5 & MP & 137.1 & 262.4 & 1.91 & $1.86 \times 10^{48}$ & $3.64 \times 10^{49}$ & $6.04 \times 10^{49}$ \\
B10\_Omg0 & $10^{10}$ & 0 & BH & 646.3 & 53.2 & 2.49 & $6.34 \times 10^{44}$ & $4.23 \times 10^{48}$ & -- \\
B10\_Omg01 & $10^{10}$ & 0.1 & BP & 373.0 & 944.0 & 2.26 & $7.68 \times 10^{47}$ & $1.25 \times 10^{50}$ & $4.04 \times 10^{50}$ \\
B10\_Omg02 & $10^{10}$ & 0.2 & MP & 506.6 & 386.1 & 2.40 & $2.50 \times 10^{47}$ & $1.69 \times 10^{49}$ & $3.97 \times 10^{49}$ \\
B10\_Omg03 & $10^{10}$ & 0.3 & MP & 430.1 & 938.6 & 2.32 & $5.43 \times 10^{47}$ & $7.61 \times 10^{49}$ & $3.46 \times 10^{50}$ \\
B10\_Omg04 & $10^{10}$ & 0.4 & BP & 355.5 & 449.7 & 2.26 & $3.02 \times 10^{47}$ & $5.78 \times 10^{49}$ & $1.74 \times 10^{50}$ \\
B10\_Omg05 & $10^{10}$ & 0.5 & BP & 110.7 & 328.0 & 1.83 & $4.42 \times 10^{48}$ & $3.99 \times 10^{49}$ & $8.13 \times 10^{49}$ \\
B11\_Omg0 & $10^{11}$ & 0 & BH & 659.2 & 45.1 & 2.50 & $7.88 \times 10^{44}$ & $3.90 \times 10^{48}$ & -- \\
B2e11\_Omg0 & $2 \times 10^{11}$ & 0 & BH & 652.3 & 47.1 & 2.50 & $2.59 \times 10^{46}$ & $4.23 \times 10^{48}$ & -- \\
B4e11\_Omg0 & $4 \times 10^{11}$ & 0 & ND & 560.2 & 750.0 & 2.41 & $1.23 \times 10^{49}$ & $9.74 \times 10^{49}$ & $5.12 \times 10^{50}$ \\
B6e11\_Omg0 & $6 \times 10^{11}$ & 0 & ND & 512.8 & 639.1 & 2.38 & $1.62 \times 10^{49}$ & $8.53 \times 10^{49}$ & $4.95 \times 10^{50}$ \\
B8e11\_Omg0 & $8 \times 10^{11}$ & 0 & ND & 499.8 & 653.5 & 2.37 & $2.87 \times 10^{49}$ & $8.72 \times 10^{49}$ & $5.19 \times 10^{50}$ \\
B11\_Omg01 & $10^{11}$ & 0.1 & MP & 503.6 & 554.6 & 2.39 & $2.64 \times 10^{48}$ & $5.40 \times 10^{49}$ & $1.57 \times 10^{50}$ \\
B11\_Omg02 & $10^{11}$ & 0.2 & BP & 294.4 & 649.4 & 2.18 & $4.01 \times 10^{48}$ & $5.25 \times 10^{49}$ & $2.79 \times 10^{50}$ \\
B11\_Omg03 & $10^{11}$ & 0.3 & BP & 216.3 & 439.5 & 2.08 & $3.45 \times 10^{48}$ & $3.00 \times 10^{49}$ & $1.20 \times 10^{50}$ \\
B11\_Omg04 & $10^{11}$ & 0.4 & BP & 189.2 & 527.3 & 2.02 & $5.04 \times 10^{48}$ & $3.89 \times 10^{49}$ & $1.55 \times 10^{50}$ \\
B11\_Omg05 & $10^{11}$ & 0.5 & BP & 193.0 & 462.3 & 2.04 & $4.32 \times 10^{48}$ & $3.45 \times 10^{49}$ & $1.15 \times 10^{50}$ \\
B12\_Omg0 & $10^{12}$ & 0 & ND & 370.9 & 620.5 & 2.24 & $4.12 \times 10^{49}$ & $1.18 \times 10^{50}$ & $7.16 \times 10^{50}$ \\
B12\_Omg01 & $10^{12}$ & 0.1 & ND & 303.6 & 838.9 & 2.15 & $4.78 \times 10^{49}$ & $1.04 \times 10^{50}$ & $8.76 \times 10^{50}$ \\
B12\_Omg02 & $10^{12}$ & 0.2 & ND & 252.6 & 938.1 & 2.04 & $5.28 \times 10^{49}$ & $8.45 \times 10^{49}$ & $1.07 \times 10^{51}$ \\
B12\_Omg03 & $10^{12}$ & 0.3 & ND & 164.9 & 625.6 & 1.91 & $8.40 \times 10^{49}$ & $1.00 \times 10^{50}$ & $7.76 \times 10^{50}$ \\
B12\_Omg04 & $10^{12}$ & 0.4 & BP & 153.6 & 620.1 & 1.87 & $7.36 \times 10^{49}$ & $9.29 \times 10^{49}$ & $8.93 \times 10^{50}$ \\
B12\_Omg05 & $10^{12}$ & 0.5 & BP & 103.8 & 523.2 & 1.75 & $1.03 \times 10^{50}$ & $1.10 \times 10^{50}$ & $6.40 \times 10^{50}$ \\
\enddata
\vspace{-0.8cm}
\tablecomments{Model parameters and simulation results. Columns represent: (1) Model Name, (2) Initial magnetic field strength $B_0$ in code units, corresponding to $\sqrt{4\pi}$~Gauss, (3) Initial angular velocity $\Omega_0$, (4) Final morphology, (5) End time of simulation, (6) Shock radius at end time, (7) Proto-neutron star mass at end time, (8) Magnetic energy, (9) Kinetic energy, and (10) Diagnostic explosion energy.}
\end{deluxetable}

\section{Simulation overview} \label{sec:overview}

\subsection{Non-rotating models}

\begin{figure}[]
    \centering
    \includegraphics[width=0.45\textwidth]{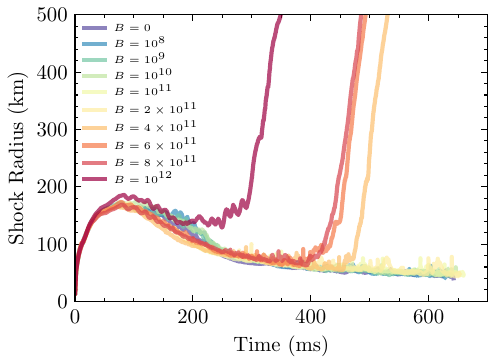}
    \includegraphics[width=0.45\textwidth]{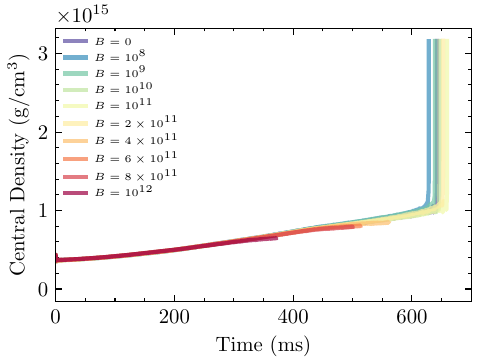}
    \includegraphics[width=0.45\textwidth]{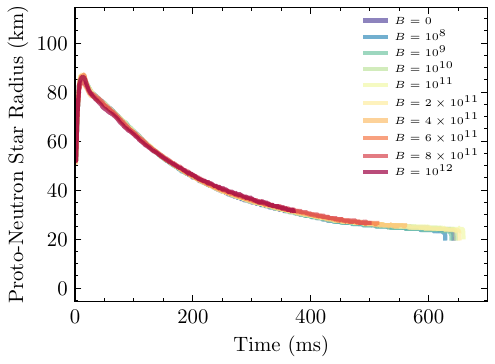}
    \includegraphics[width=0.45\textwidth]{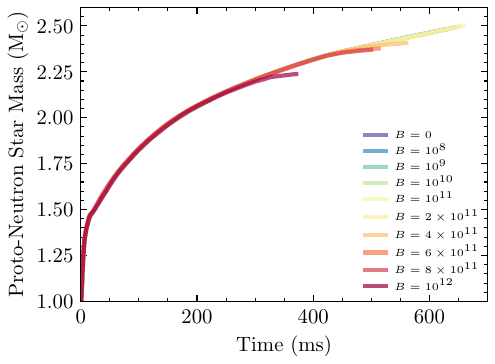}
    \caption{\label{fig:non_rotating}
    Time evolution of the averaged shock radius (upper left), central density (upper right), proto-neutron star radius (lower left), and proto-neutron star mass (lower right) for non-rotating models ($\Omega_0=0$). Different colors represent models with different initial magnetic field strengths ranging from $B_0=0$ to $10^{12}$ (see Table~\ref{tab:simulations}). All non-rotating models with $B_0 \leq 2 \times 10^{11}$ result in failed explosions and black hole formation, while models with $B_0 \geq 4 \times 10^{11}$ exhibit successful neutrino-driven explosions.}
\end{figure}

To isolate the effects of magnetic fields on the explosion dynamics, we begin with the non-rotating models ($\Omega_0=0$). Figure~\ref{fig:non_rotating} presents the temporal evolution of key physical quantities for initial magnetic field strengths ranging from $B_0=0$ to $10^{12}$.

The shock radius evolution (upper left panel) reveals a critical magnetic field threshold: models with $B_0 \leq 2 \times 10^{11}$ exhibit shock stagnation followed by contraction, ultimately leading to black hole formation, whereas models with $B_0 \geq 4 \times 10^{11}$ show sustained shock expansion driven by strong magnetic pressure and neutrino heating. The central density (upper right panel) rises to nuclear values at core bounce ($\rho_c \sim 3 \times 10^{14}$~g~cm$^{-3}$); in the failed models it continues to increase until the rapid rise signals black hole formation. The PNS radius (lower left panel) evolves similarly across all models prior to the bifurcation; the key difference is that exploding models are terminated shortly after shock revival, while non-exploding models are followed to longer times until black hole formation. The PNS mass (lower right panel) likewise shows little variation before explosion. In the successful models, the reduced mass accretion rate following shock revival leads to a slower growth of the PNS mass, whereas the failed models continue to accrete at higher rates, driving the PNS mass toward the maximum gravitational mass of $\sim 2.5~M_\odot$ for the SFHo equation of state.

\subsection{Rotating models}

\begin{figure}[]
    \centering
    \includegraphics[width=0.45\textwidth]{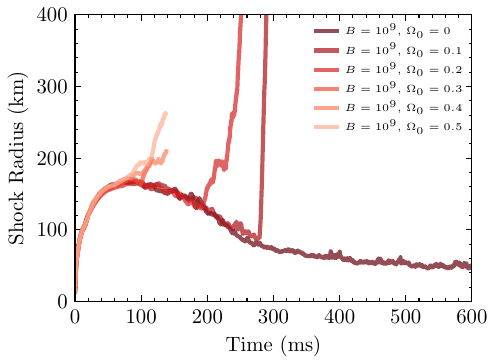}
    \includegraphics[width=0.45\textwidth]{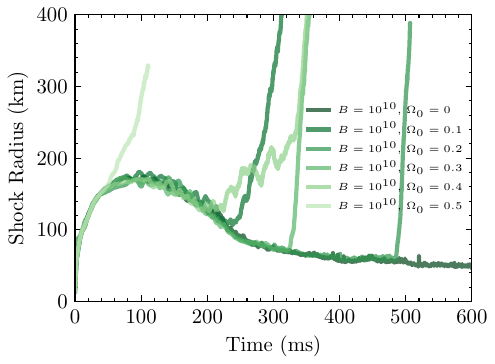}
    \includegraphics[width=0.45\textwidth]{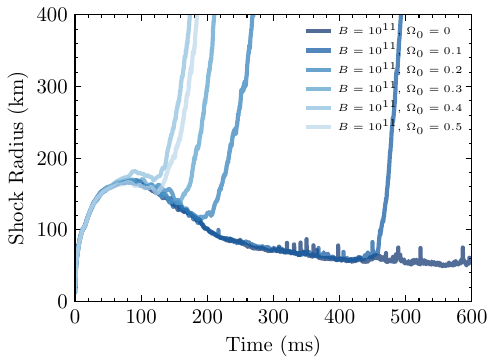}
    \includegraphics[width=0.45\textwidth]{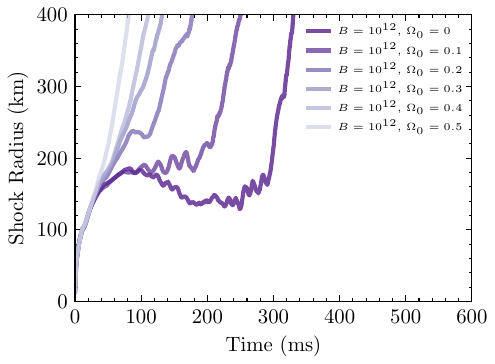}
    \caption{\label{fig:rotating_rshock}
    Time evolution of the averaged shock radius for rotating models with different initial magnetic field strengths: $B_0=10^9$ (upper left), $10^{10}$ (upper right), $10^{11}$ (lower left), and $10^{12}$ (lower right). Different transparencies represent models with different initial angular velocities ranging from $\Omega_0=0.1$ to $0.5$~rad~s$^{-1}$. The shock evolution demonstrates strong dependence on both magnetic field strength and rotation rate, with higher values generally leading to faster shock expansion and earlier explosions.}
\end{figure}

 Rotation fundamentally alters the explosion landscape by enabling magnetorotational mechanisms and modifying the neutrino-driven dynamics. The shock radius evolution for rotating models (Figure~\ref{fig:rotating_rshock}) illustrates this across four magnetic field regimes. At weak fields ($B_0=10^9$, upper left), all rotating models achieve successful shock revival, with faster rotation rates leading to earlier onset of explosion. At intermediate field strengths ($B_0=10^{10}$ and $10^{11}$, upper right and lower left), the interplay between magnetic fields and rotation becomes more complex. While all rotating models ultimately achieve shock revival, the onset of explosion does not follow a simple monotonic trend with rotation rate, suggesting nonlinear interactions between the magnetorotational and neutrino-driven mechanisms. For the strongest field ($B_0=10^{12}$, lower right), all rotating models explode rapidly ($<300$~ms), demonstrating the dominant role of magnetic pressure when combined with rotational support. Throughout, the shock expansion timescale decreases systematically with increasing rotation rate, reflecting enhanced angular momentum transport and magnetic field amplification.

\begin{figure}[]
    \centering
    \includegraphics[width=0.45\textwidth]{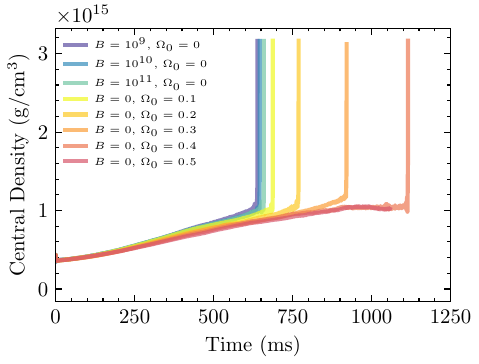}
    \caption{\label{fig:rotating_bh}
    Time evolution of central density for all models that result in black hole formation. Different colors represent models with different magnetic field strengths and rotational speeds (see Table~\ref{tab:simulations}). The sharp increment in central density marks the moment of black hole formation. All non-rotating models without sufficiently strong magnetic fields ($B_0 < 4 \times 10^{11}$) undergo black hole formation, while some rotating models also fail to explode despite the presence of rotation and magnetic fields.}
\end{figure}

Not all models succeed in exploding. The central density evolution for all failed models (Figure~\ref{fig:rotating_bh}) traces the transition from proto-neutron star to black hole, marked by an abrupt density increase as matter collapses within the apparent horizon. Black hole formation occurs in two categories of models: (1) all non-rotating models without magnetic fields ($B_0=0$), which inevitably collapse within 600--1100~ms post-bounce, and (2) non-rotating models with weak-to-intermediate magnetic fields ($B_0 \leq 2 \times 10^{11}$), where the magnetic pressure alone is insufficient to drive an explosion. Notably, the inclusion of even modest rotation universally prevents black hole formation in magnetized models, underscoring the critical role of rotation in activating the magnetorotational mechanism for this massive progenitor.

\begin{figure}[]
    \centering
    \includegraphics[width=0.45\textwidth]{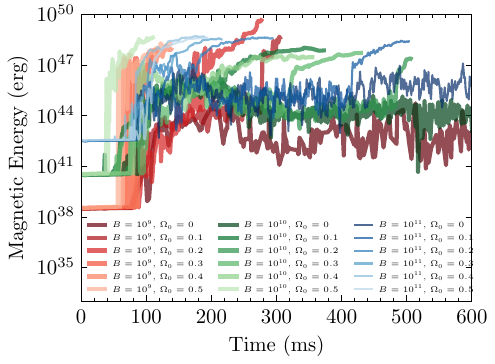}
    \includegraphics[width=0.45\textwidth]{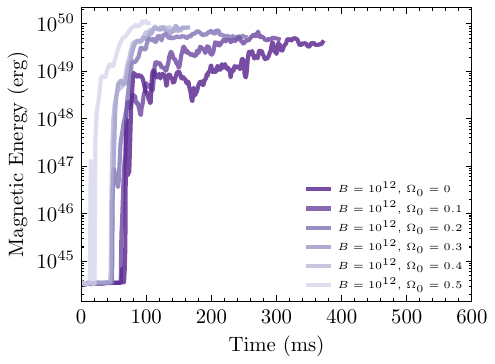}
    \caption{\label{fig:rotating_emag}
    Time evolution of magnetic energy for rotating models with $B_0=10^9$, $10^{10}$, and $10^{11}$ (left) and $B_0=10^{12}$ (right). Different transparencies and colors represent models with different initial angular velocities and magnetic field strengths. The magnetic energy shows rapid amplification in successfully exploding models, reaching values several orders of magnitude higher than the initial magnetic energy, indicating efficient magnetic field amplification via MHD instabilities and rotational winding in the post-shock region.}
\end{figure}

Successful explosions are accompanied by dramatic magnetic field amplification, as shown by the magnetic energy evolution in Figure~\ref{fig:rotating_emag}. For initial strengths $B_0=10^9$--$10^{11}$ (left panel), the magnetic energy grows by 7--11 orders of magnitude, driven by flux compression during collapse, magnetic winding of the differential rotation, and the magnetorotational instability (MRI). At the highest field strength ($B_0=10^{12}$, right panel), the evolution is instead dominated by the strong initial configuration, with rapid magnetic winding generating dynamically important toroidal fields promptly after bounce. The magnetic energy saturates as magnetic pressure approaches equipartition with thermal and ram pressures, consistent with the saturation behavior observed in previous MHD studies \cite{2015Natur.528..376M,2016MNRAS.460.3316R}. Across all successful models, the final magnetic energies reach $\sim 10^{48}$--$10^{50}$~erg, which are sufficient to power relativistic jets in the most magnetized, rapidly rotating cases.

\begin{figure}[]
    \centering
    \includegraphics[width=0.45\textwidth]{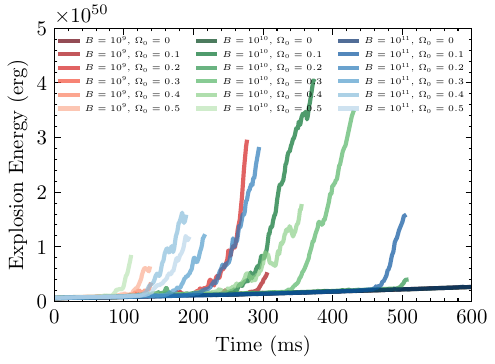}
    \includegraphics[width=0.45\textwidth]{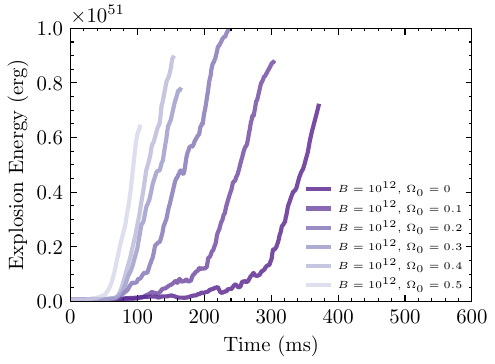}
    \caption{\label{fig:rotating_edia}
    Time evolution of diagnostic explosion energy for rotating models with $B_0=10^9$, $10^{10}$, and $10^{11}$ (left) and $B_0=10^{12}$ (right). Different transparencies and colors represent models with different initial angular velocities and magnetic field strengths. Models that achieve positive and sustained growth in diagnostic energy are classified as successful explosions, while those showing stagnation or decline lead to failed explosions.}
\end{figure}

The energetics of these explosions are quantified by the diagnostic explosion energy (Figure~\ref{fig:rotating_edia}). 
At weaker magnetization ($B_0=10^9$--$10^{11}$, left panel), explosion energies span $10^{49}$--$10^{50}$~erg, comparable to typical core-collapse supernovae, with strong dependence on both rotation rate and field strength. The most energetic explosions occur in strongly magnetized models ($B_0=10^{12}$, right panel), reaching $E_{\rm dia} \sim 10^{51}$~erg within a few hundreds of miliseoncds, making them potential hypernovae and long gamma-ray burst progenitors.

\begin{figure}[]
    \centering
    \includegraphics[width=0.6\textwidth]{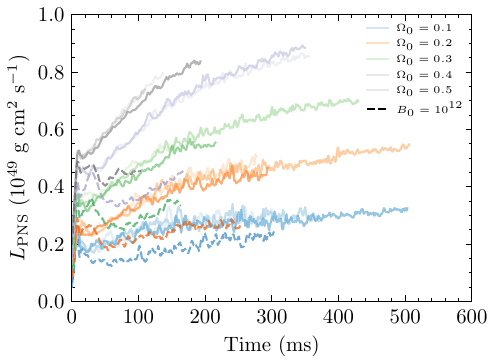}
    \caption{\label{fig:lpns_evolution}
    Time evolution of the proto-neutron star angular momentum with models varying rotation rates from $\Omega_0=0.1$ to $0.5$~rad~s$^{-1}$. Different colors represent models with different inital rotational rates and solid lines represent models with different initial magnetic filed strengths ($B_0=10^{9}, 10^{10}, 10^{11}$), while dashed lines indicate the models with initial magnetic field strength $B_0=10^{12}$. The PNS angular momentum evolution reflects the competition between angular momentum transport via magnetic braking and magnetohydrodynamic torques, and angular momentum conservation during mass ejection. Models with higher initial rotation rates maintain larger PNS angular momenta throughout the evolution, with values reaching $\sim 0.3$--$0.9 \times 10^{49}$~g~cm$^2$~s$^{-1}$ at late times. The sustained high angular momentum in rapidly rotating models is crucial for maintaining centrifugal support and enabling the formation of collimated bipolar jets characteristic of magnetorotational explosions.}
\end{figure}

Further insight into the role of rotation comes from the PNS angular momentum evolution (Figure~\ref{fig:lpns_evolution}). Models with higher initial rotation rates maintain larger PNS angular momenta throughout the post-bounce phase, reaching $\sim 0.3$--$0.9 \times 10^{49}$~g~cm$^2$~s$^{-1}$ at late times. This evolution reflects a competition between angular momentum transport via magnetic braking and MHD torques on one hand, and angular momentum conservation during accretion and ejection on the other. Notably, the strongest-field models ($B_0 = 10^{12}$, dashed lines) exhibit more efficient angular momentum extraction than their weaker-field counterparts (solid lines). The sustained high angular momentum in rapidly rotating models is crucial for maintaining centrifugal support and enabling the formation of collimated bipolar jets.

\section{Ejecta Morphology} \label{sec:explosion}

\begin{figure}[]
    \centering
    \includegraphics[width=1.0\textwidth]{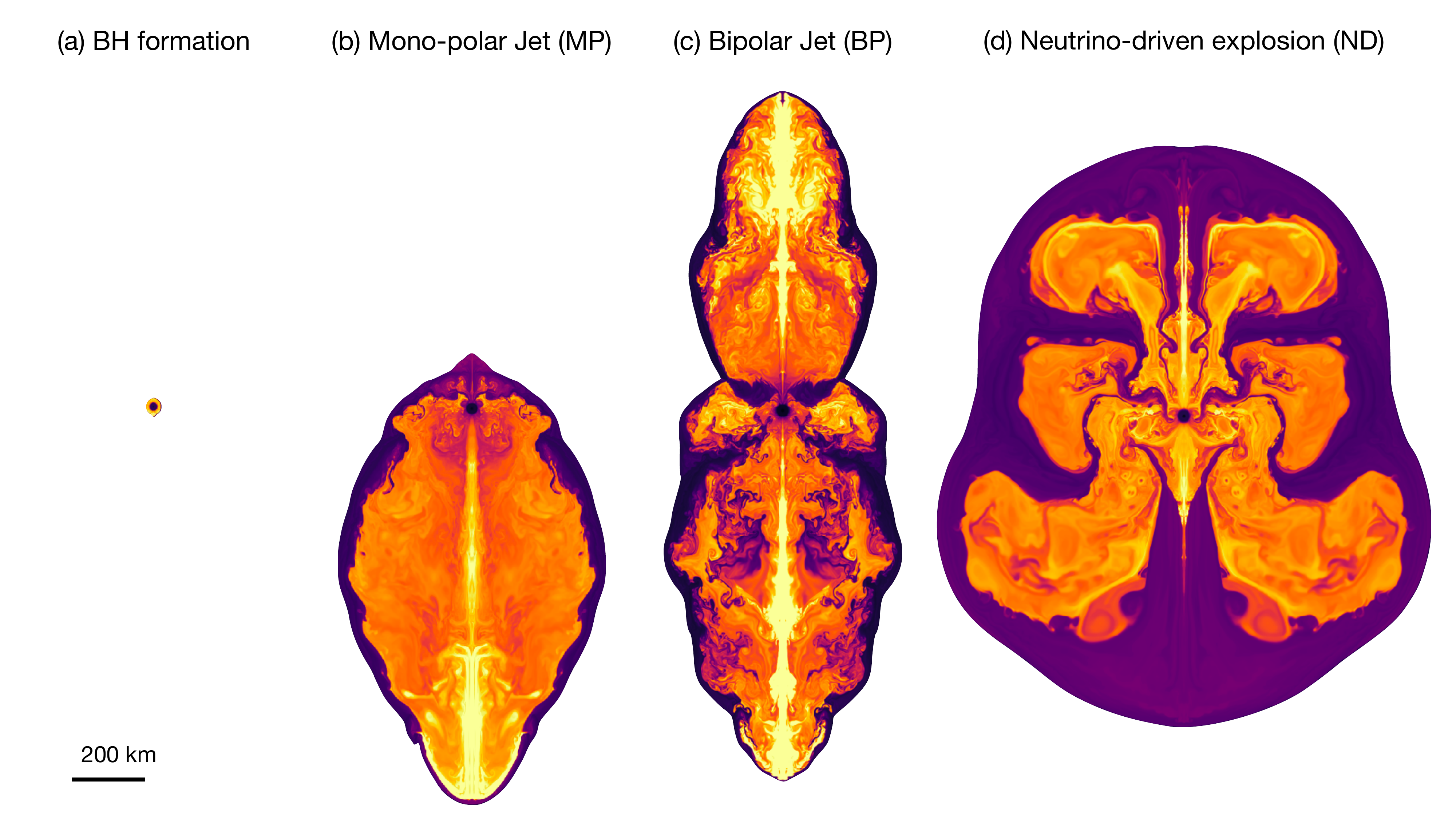}
    \caption{\label{fig:morphology}
    Entropy distribution snapshots illustrating the four distinct explosion morphologies identified in our simulations: (a) failed supernova leading to black hole formation (model B0\_Omg05), (b) monopolar jet explosion (model B9\_Omg01), (c) bipolar jet explosion (model B12\_Omg05), and (d) neutrino-driven explosion (model B12\_Omg02). The color scale represents specific entropy in units of $k_{\rm B}$ baryon$^{-1}$. Different morphologies result from the interplay between magnetic field strength, rotation rate, and neutrino heating, leading to diverse explosion mechanisms and ejecta geometries.}
\end{figure}

The interplay between magnetic fields, rotation, and neutrino heating gives rise to four distinct explosion morphologies, illustrated by representative entropy snapshots in Figure~\ref{fig:morphology}. Failed explosions (panel a, model B0\_Omg05) exhibit weak shock expansion and eventual collapse to a black hole, characterized by high-entropy ($s > 30~k_{\rm B}$ baryon$^{-1}$) stagnated material surrounding the collapsing PNS. Monopolar jet explosions (panel b, model B9\_Omg01) produce asymmetric outflows preferentially ejected along one pole, arising from slight initial asymmetries amplified by MHD and SASI instabilities in moderately magnetized, rotating systems. Bipolar jet explosions (panel c, model B12\_Omg05) generate highly collimated, symmetric jets along the rotation axis, powered by strong toroidal magnetic fields and centrifugal support; these represent the canonical magnetorotational mechanism associated with long GRB progenitors. Neutrino-driven explosions (panel d, model B12\_Omg02) display more isotropic ejecta with broader opening angles ($\sim 60^\circ$--$90^\circ$), where strong neutrino heating drives shock revival aided by, but not dominated by, magnetic pressure.

\begin{figure}[]
    \centering
    \includegraphics[width=1.0\textwidth]{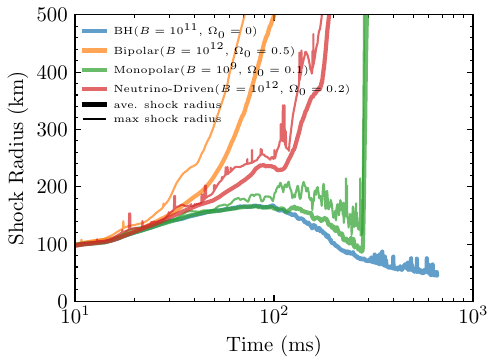}
    \caption{\label{fig:morphology_rshock}
    Time evolution of the averaged (thick lines) and maximum (thin lines) shock radii for representative models of each morphology class. Different colors represent different explosion outcomes: black hole formation (blue), monopolar jet (green), bipolar jet (orange), and neutrino-driven explosion (red). The divergence between averaged and maximum shock radii indicates the degree of asphericity in the explosion, with jet-driven explosions (monopolar and bipolar) showing the largest deviations due to highly anisotropic ejecta morphologies.}
\end{figure}

These morphological classes are quantitatively distinguished by the shock radius evolution (Figure~\ref{fig:morphology_rshock}), where the divergence between the averaged (thick lines) and maximum (thin lines) shock radii serves as a direct measure of explosion asphericity. Black hole formation models (blue) show minimal divergence as the shock stagnates at small radii ($\lesssim 100$~km) before collapse. Neutrino-driven explosions (red) exhibit moderate asphericity, reflecting large-scale convective and SASI-driven asymmetries. Monopolar (green) and bipolar (orange) jet explosions display the most extreme asphericity, with maximum-to-average ratios reaching $5$--$10$ as collimated jets propagate rapidly along the polar axis while equatorial regions remain confined.

\begin{figure}[]
    \centering
    \includegraphics[width=0.45\textwidth]{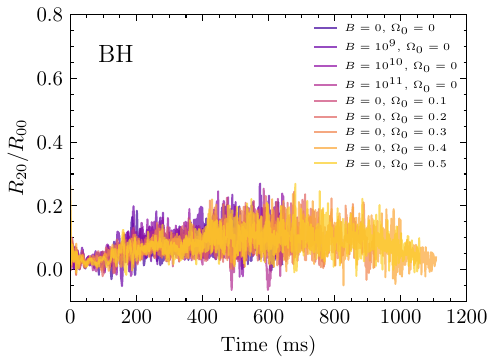}
    \includegraphics[width=0.45\textwidth]{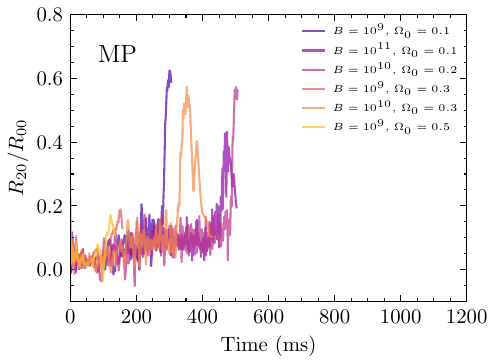}
    \includegraphics[width=0.45\textwidth]{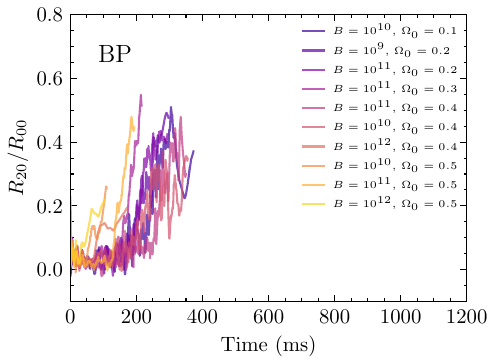}
    \includegraphics[width=0.45\textwidth]{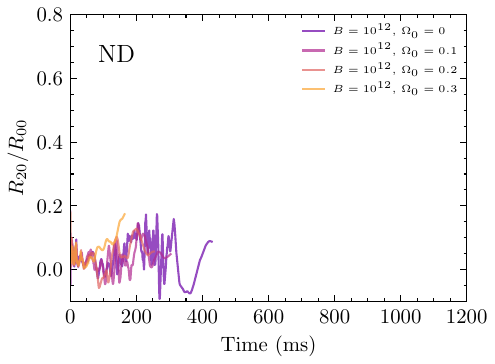}
    \caption{\label{fig:asymmetry_r20r00}
    Time evolution of the asymmetry parameter $R_{20}/R_{00}$ derived from spherical harmonic decomposition for models of each morphology class: Black hole formation (BH; upper left), Monopolar jet explosions (MP; upper right), Bipolar jet explosions (BP; lower left), and Neutrino-driven explosions (ND; lower right). Different colors represent different models within each morphology class. The ratio $R_{20}/R_{00}$ quantifies the degree of asphericity in the mass distribution, with higher values indicating stronger quadrupole deformations. Black hole formation models maintain low asymmetry ($R_{20}/R_{00} \lesssim 0.2$) throughout their evolution, while jet-driven explosions (monopolar and bipolar) exhibit pronounced asymmetries with $R_{20}/R_{00}$ reaching $0.4$--$0.6$ during active jet propagation, reflecting the highly anisotropic nature of magnetorotational explosions.}
\end{figure}

A more rigorous characterization of the asphericity is obtained by decomposing the shock surface into spherical harmonics (Figure~\ref{fig:asymmetry_r20r00}). The ratio $R_{20}/R_{00}$, measuring quadrupole deformation relative to the monopole, remains low ($\lesssim 0.2$) for both BH formation models (upper left) and neutrino-driven explosions (lower right), though the two cases differ in their temporal behavior: BH formation models exhibit higher-frequency oscillations in $R_{20}/R_{00}$, indicative of vigorous SASI activity behind the stagnating shock, whereas the neutrino-driven cases show smoother evolution consistent with their broader, more isotropic expansion. Monopolar and bipolar jet explosions (upper right and lower left) develop pronounced deformations with $R_{20}/R_{00}$ reaching $0.4$--$0.6$ during active jet propagation, reflecting the highly elongated, pole-dominated ejecta geometry. This ratio provides a diagnostic complementary to the shock radius analysis and is directly connected to the gravitational wave emission discussed in Section~\ref{sec:gw}.

\begin{figure}[]
    \centering
    \includegraphics[width=0.4\textwidth]{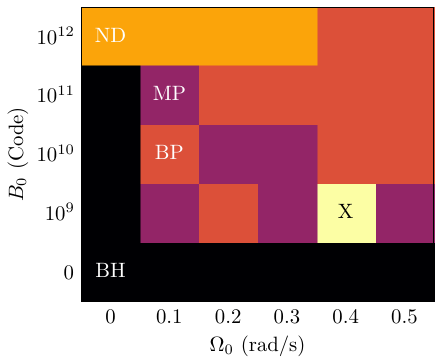}
    \includegraphics[width=0.5\textwidth]{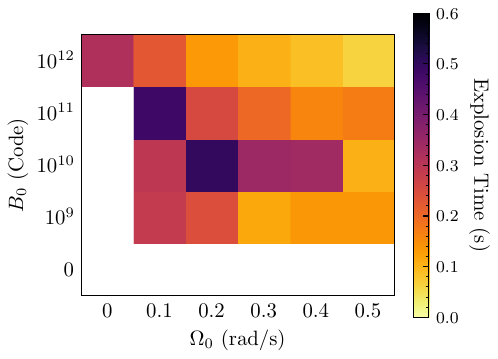}
    \caption{\label{fig:state}
    Left: Morphology classification in the $B_0$-$\Omega_0$ parameter space. Different colors represent different explosion outcomes: black (BH) for failed explosions leading to black hole formation, purple (MP) for monopolar jet explosions, red (BP) for bipolar jet explosions, orange (ND) for neutrino-driven explosions, and yellow (X) for an unclassified case. Right: Explosion time as a function of initial magnetic field strength and rotation rate. The color scale indicates the explosion time in seconds. White regions correspond to models that fail to explode. The results demonstrate that stronger magnetic fields and faster rotation generally lead to earlier explosions, with the shortest explosion times occurring at high $B_0$ and high $\Omega_0$.}
\end{figure}

The full $B_0$--$\Omega_0$ parameter space is summarized in Figure~\ref{fig:state}. Clear morphological boundaries emerge (left panel): failed explosions (black, BH) dominate the low-magnetization, low-rotation regime where neither neutrino heating nor magnetorotational mechanisms provide sufficient energy to unbind the envelope. Neutrino-driven explosions (orange, ND) occupy a narrow region at very high magnetic fields ($B_0 \gtrsim 4 \times 10^{11}$) with weak-to-moderate rotation. Bipolar jets (red, BP) appear at high $B_0$ and high $\Omega_0$, representing the magnetorotational-dominated regime, while monopolar jets (purple, MP) fill the transitional zone at intermediate parameters. The explosion timescale (right panel) decreases systematically with both increasing field strength and rotation rate, from $>500$~ms in marginally successful models to $<150$~ms in strongly magnetized, rapidly rotating systems, reflecting the efficiency of magnetorotational energy extraction in accelerating shock revival.

\section{Gravitational Wave Emissions} \label{sec:gw}

The diverse explosion dynamics described above produce distinctive gravitational wave (GW) signatures that encode information about the explosion mechanism, progenitor rotation, and magnetic field strength. As noted in Section~\ref{sec:methods}, our axisymmetric simulations capture only the $m=0$ component of the GW emission; non-axisymmetric contributions from spiral SASI modes, bar-mode instabilities, and low-$T/|W|$ instabilities, which can dominate the signal in rapidly rotating 3D models \cite{2024MNRAS.531.3732S}, are absent by construction. The amplitudes and spectral content presented here should therefore be interpreted as partial views of the full three-dimensional signal.

\begin{figure}[]
    \centering
    \includegraphics[width=0.45\textwidth]{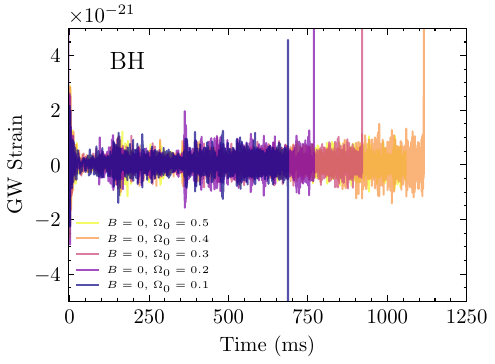}
    \includegraphics[width=0.45\textwidth]{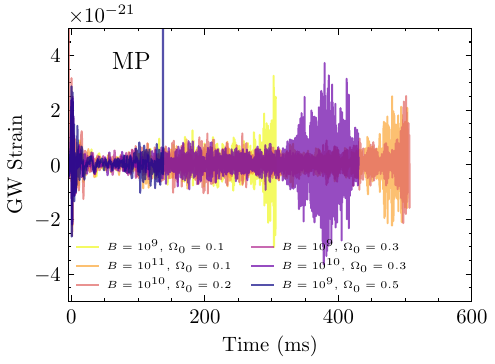}
    \includegraphics[width=0.45\textwidth]{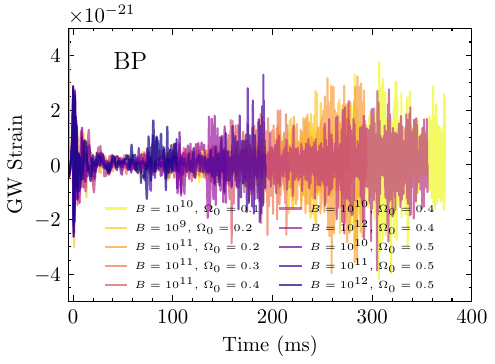}
    \includegraphics[width=0.45\textwidth]{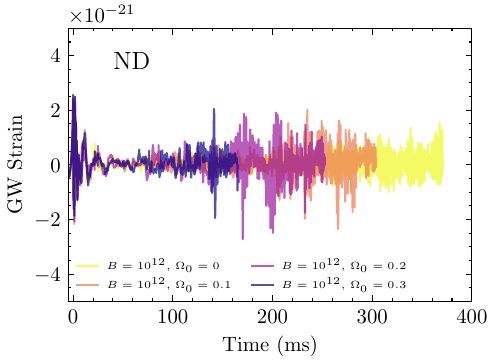}
    \caption{\label{fig:gw_strain}
    Time-domain gravitational wave strain waveforms for models resulting in different explosion morphologies at a fiducial distance of $10$~kpc: Black hole formation models (BH; upper left), Monopolar jet explosions (MP; upper right), Bipolar jet explosions (BP; lower left), and Neutrino-driven explosions (ND; lower right). Different colors represent different models within each morphology class (see Table~\ref{tab:simulations}). The initial spike corresponds to core bounce, while the subsequent evolution reflects proto-neutron star oscillations, convective instabilities, and asymmetric mass motions characteristic of each explosion mechanism.}
\end{figure}

The GW strain waveforms for all four morphological classes are shown in Figure~\ref{fig:gw_strain} at a fiducial distance of 10~kpc. BH formation models (upper left) display a sharp bounce signal at $t \approx 0$ followed by oscillations from PNS convection and SASI activity that maintain characteristic strain amplitudes of $h \sim 10^{-21}$, with the GW frequency progressively increasing over time as the PNS contracts and becomes more compact before the apparent horizon forms. Monopolar and bipolar jet explosions (upper right and lower left) exhibit pre-explosion signals similar to the BH formation cases, but once the explosion is launched, they produce significantly enhanced GW emission with $h \sim 2$--$4 \times 10^{-21}$, driven by asymmetric mass ejection and, in the bipolar case, quasi-periodic modulations from rapid PNS rotation and rotating quadrupole deformations. Neutrino-driven explosions (lower right) generate GW amplitudes comparable to the BH formation models ($h \sim 10^{-21}$), but at characteristically lower frequencies, reflecting the slower, more stochastic mass motions dominated by convective overturn and SASI oscillations rather than the high-frequency PNS oscillations seen in the collapsing BH cases.

\begin{figure}[]
    \centering
    \includegraphics[width=0.45\textwidth]{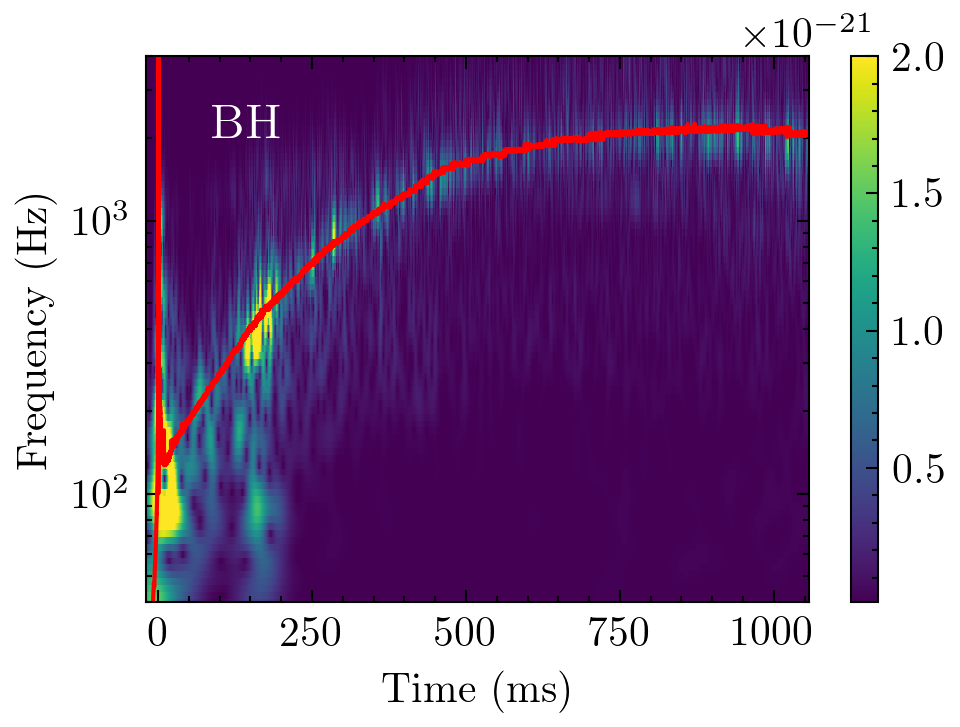}
    \includegraphics[width=0.45\textwidth]{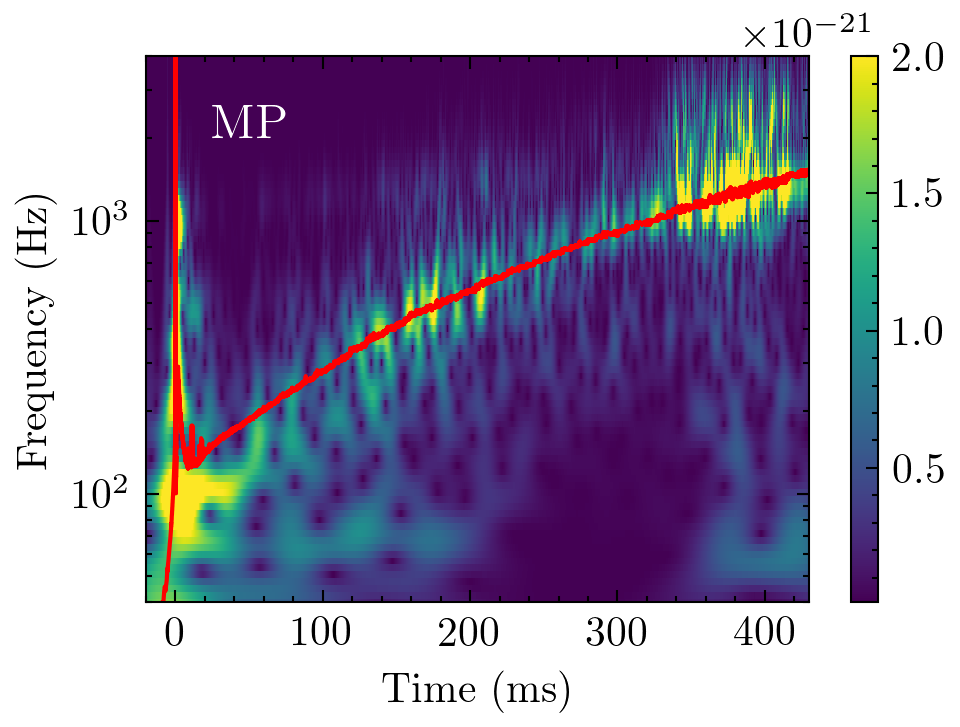}
    \includegraphics[width=0.45\textwidth]{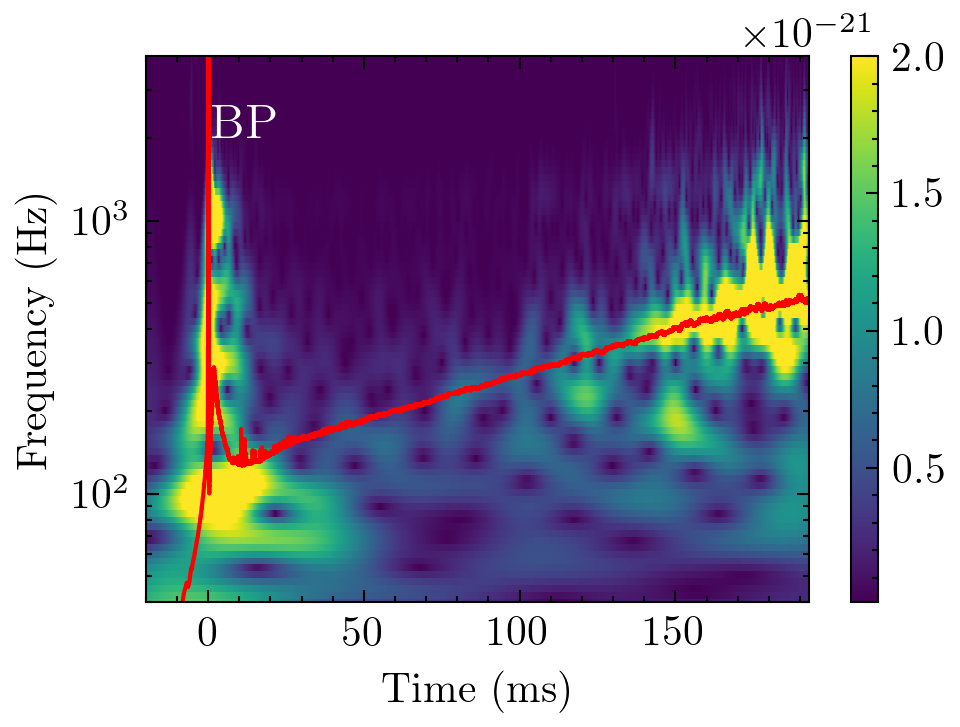}
    \includegraphics[width=0.45\textwidth]{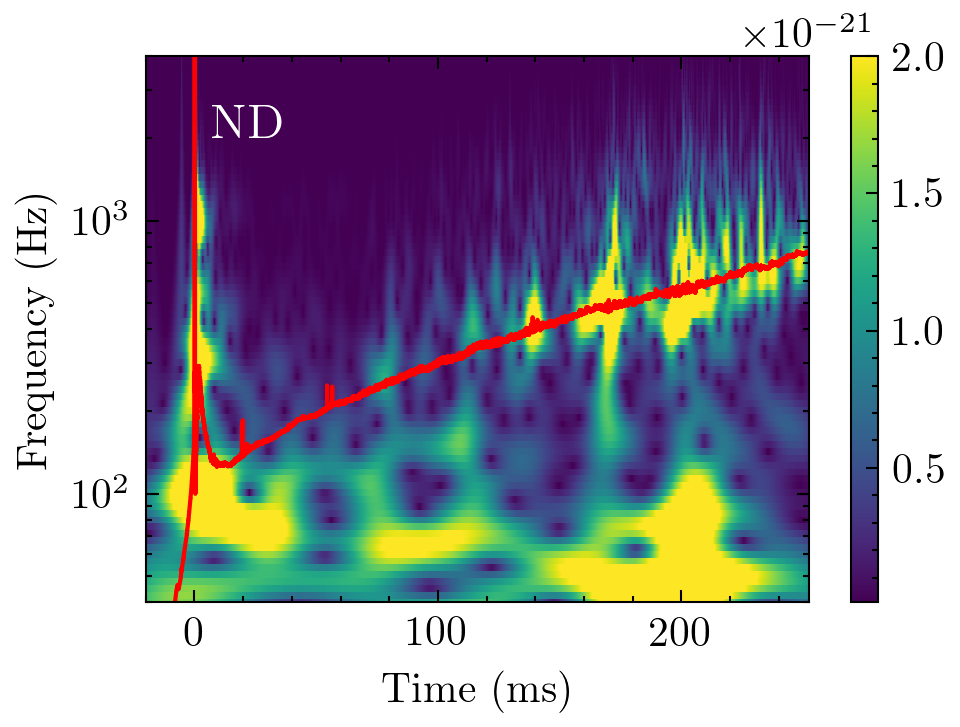}
    \caption{\label{fig:gw_spectrogram}
    Gravitational wave spectrograms for representative models of each morphology class: Black hole formation (model B0\_Omg05; upper left), Mnopolar jet explosion (model B10\_Omg03; upper right), Bipolar jet explosion (model B11\_Omg05; lower left), and Neutrino-driven explosion (model B12\_Omg02; lower right). The red solid lines track the dominant GW peak frequency evolution. The spectrograms reveal distinct frequency evolution patterns for different explosion mechanisms, with rotating models showing monotonically increasing frequencies due to proto-neutron star contraction and spin-up, while the failed explosion model exhibits a plateau at late times as the PNS approaches black hole formation.}
\end{figure}

Time-frequency analysis via spectrograms further reveals the spectral evolution of GW emission across the different explosion mechanisms (Figure~\ref{fig:gw_spectrogram}). The failed explosion model (upper left, B0\_Omg05) shows initial broadband emission at $\sim 200$--$800$~Hz from bounce and early convection, followed by the fundamental mode of the PNS oscillation, ranging from $\sim$ hundred Hz to kilo-Hz when approaches black hole formation. The monopolar jet model (upper right, B10\_Omg03) and the bipolar jet model (lower left, B11\_Omg05) exhibit a similar upward frequency drift from $\sim 200$~Hz to $>1000$~Hz, driven by PNS contraction and spin-up as angular momentum is conserved during mass loss. The neutrino-driven model (lower right, B12\_Omg02) shows weaker rotation signatures, with more diffuse spectral power concentrated below $600$~Hz, reflecting the dominance of convective rather than rotational dynamics.

\begin{figure}[]
    \centering
    \includegraphics[width=1\textwidth]{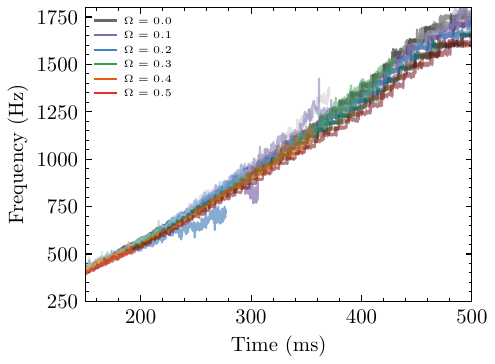}
    \caption{\label{fig:gw_fpeak}
    Time evolution of the dominant gravitational wave peak frequency for all models. Different colors represent models with different initial angular velocities ranging from $\Omega_0=0$ to $0.5$~rad~s$^{-1}$. The GW peak frequency increases monotonically with time for all models due to proto-neutron star contraction and spin-up, and is not sensitive to the magnetic field strength. Models with higher initial rotation rates exhibit lower peak frequency evolution. }
\end{figure}

Compiling the peak GW frequency for all 34 models (Figure~\ref{fig:gw_fpeak}) reveals that $f_{\rm peak}$ is primarily governed by PNS contraction and is far more sensitive to the initial rotation rate than to the magnetic field strength.
All models exhibit a monotonically increasing $f_{\rm peak}$, rising from $\sim 250$--$500$~Hz shortly after bounce to $\sim 1500$--$1750$~Hz at late times, consistent with the fundamental ($f$-) mode of the contracting PNS \cite{2019MNRAS.482.3967T,2024ApJ...961..194H}.
The most striking feature is that models sharing the same $\Omega_0$ but spanning a wide range of $B_0$ produce nearly identical $f_{\rm peak}$ tracks, demonstrating that the GW peak frequency is insensitive to the magnetic field strength over the parameter range explored.
In contrast, different rotation rates introduce a modest but systematic spread: models with higher $\Omega_0$ show slightly lower $f_{\rm peak}$ at a given post-bounce time, reflecting the centrifugal support that slows PNS contraction and correspondingly reduces the oscillation frequency.
This rotation-dependent grouping suggests that the GW peak frequency can serve as a diagnostic of progenitor core rotation, largely independent of the magnetic field configuration \cite{2019ApJ...878...13P,2021ApJ...914...80P}.
Our findings are qualitatively consistent with recent 3D GRMHD simulations by \cite{2024MNRAS.531.3732S}, who find similarly monotonic $f_{\rm peak}$ increases in rapidly rotating, magnetized models of a 20~$M_\odot$ progenitor with spectral neutrino transport.

\begin{figure}[]
    \centering
    \includegraphics[width=1\textwidth]{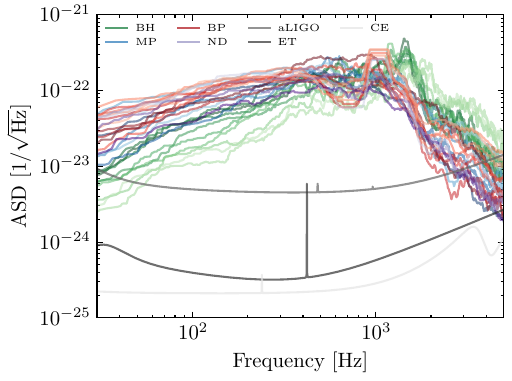}
    \caption{\label{fig:gw_asd}
    Gravitational wave amplitude spectral density (ASD) for all simulated models, assuming a source distance of $10$~kpc. Different colors represent models with different explosion morphologies: green (BH) for black hole formation, blue (MP) for monopolar jets, red (BP) for bipolar jets, and purple (ND) for neutrino-driven explosions. The gray lines show the designed sensitivity curves for advanced LIGO (aLIGO), Einstein Telescope (ET), and Cosmic Explorer (CE). Most of our models produce GW signals in the frequency range of $\sim 100$--$2000$~Hz with characteristic strain amplitudes of $\sim 10^{-23}$--$10^{-22}$~Hz$^{-1/2}$, making them potentially detectable by next-generation gravitational wave detectors for Galactic sources.}
\end{figure}

Finally, the amplitude spectral density (ASD) quantifies the detectability of these signals with current and future detectors (Figure~\ref{fig:gw_asd}). At a fiducial distance of 10~kpc, our models produce characteristic strain ASDs of $\sim 10^{-23}$--$10^{-22}$~Hz$^{-1/2}$ concentrated in the $100$--$2000$~Hz band. BH formation models (green) produce weaker signals compared to the exploding models in the $\sim 100$--$600$~Hz range, but exhibit notably strong emission near $\sim 2000$~Hz, as the prolonged evolution prior to BH formation allows the PNS to cool and contract to a more compact configuration, shifting the characteristic GW frequency to higher values. All BH models remain detectable by Advanced LIGO (aLIGO; \cite{2015CQGra..32g4001L}) at 10~kpc. Monopolar jet models (blue) show broadly similar spectral morphology to the other exploding models but with relatively less power at kHz frequencies. Bipolar jets (red) and neutrino-driven explosions (purple) generate the strongest emission, with peak ASDs exceeding the aLIGO sensitivity curve by more than an order of magnitude in the $\sim 200$--$800$~Hz band. All models would be readily detected by third-generation instruments such as Einstein Telescope (ET; \cite{2011CQGra..28i4013H}) and Cosmic Explorer (CE), with detection horizons extending to several Mpc---enabling multi-messenger observations for the next Galactic or nearby extragalactic core-collapse supernova.

\section{Summary \& Conclusions} \label{sec:summary}

We have presented a systematic two-dimensional parameter-space study of magnetorotational core-collapse supernovae, exploring 34 models with varying initial magnetic field strengths ($B_0 = 0$--$10^{12}$) and rotation rates ($\Omega_0 = 0$--$0.5$~rad~s$^{-1}$) for a 40~$M_\odot$ progenitor, employing self-consistent neutrino transport via the IDSA and an effective general relativistic potential.

Our key findings are as follows. Non-rotating models with $B_0 \leq 2 \times 10^{11}$ fail to explode and form black holes within 600--1100~ms postbounce, while a threshold of $B_0 \gtrsim 4 \times 10^{11}$ is required for successful non-rotating explosions. Even modest rotation universally prevents black hole formation in magnetized models, enabling explosions at field strengths as low as $B_0 \sim 10^9$. We identify four distinct explosion morphologies: (1) failed explosions leading to black hole formation; (2) monopolar jets at intermediate fields and rotation; (3) bipolar jets at high fields and rapid rotation, representing canonical magnetorotational explosions; and (4) neutrino-driven explosions at high fields with weak-to-moderate rotation. Magnetic energy amplification spans 7--11 orders of magnitude for $B_0 = 10^9$--$10^{11}$, driven by flux compression, magnetic winding, and the MRI, with final energies reaching $\sim 10^{48}$--$10^{50}$~erg and diagnostic explosion energies approaching $\sim 10^{51}$~erg in the most energetic cases. Gravitational wave signals are detectable by Advanced LIGO for all models at 10~kpc, with bipolar jets and neutrino-driven explosions producing the strongest emission exceeding aLIGO sensitivity by more than an order of magnitude. Third-generation detectors would extend the detection horizon to several Mpc.

One should note that our simulations are performed in axisymmetry, which suppresses non-axisymmetric instabilities such as the kink instability that can disrupt jets in 3D \cite{2014ApJ...785L..29M}, though strongly magnetized jets may remain stable \cite{2018ApJ...864..171M,2021MNRAS.503.4942O}. Axisymmetry also restricts gravitational wave signals to the $m=0$ component, missing contributions from spiral SASI and bar-mode instabilities \cite{2024MNRAS.531.3732S,2020ApJ...896..102K}. Furthermore, the dynamics of shock revival and explosion energy growth can differ quantitatively between 2D and 3D \cite{2015MNRAS.453..287M}. The morphological boundaries identified in our $B_0$--$\Omega_0$ parameter space should therefore be interpreted as qualitative trends rather than precise thresholds.
Despite these caveats, our systematic 2D survey maps the landscape of magnetorotational outcomes and provides guided priors for selecting initial conditions in computationally expensive 3D studies, which will be essential to confirm the morphological boundaries and multimessenger signatures reported here.

\section*{Acknowledgments}

This work is supported by the National Science and Technology Council of Taiwan through grants 113-2112-M-007-031, 114-2112-M-007-020, and 114-2918-I-007-016, by the Center for Informatics and Computation in Astronomy (CICA) at National Tsing Hua University through a grant from the Ministry of Education of Taiwan.
KCP acknowledges Professor Carla Fr\"{o}hlich for her hospitality during his sabbatical leave at the Department of Physics and Astronomy, North Carolina State University, where part of this work was carried out.
{\tt FLASH} was in part developed by the DOE NNSA-ASC OASCR Flash Center at the University of Chicago. The simulations and data analysis have been carried out at the {\tt Taiwania-3} supercomputer in the National Center for High-Performance Computing (NCHC) in Taiwan, and on the CICA cluster at National Tsing Hua University. Analysis and visualization of simulation data were completed using the analysis toolkit {\tt yt}.

\section*{References}
\bibliographystyle{iopart-num}
\bibliography{references}

\end{document}